\documentclass[aps,pra,reprint,showpacs,amsmath,amssymb,floatfix,nofootinbib,superscriptaddress,10pt]{revtex4-1} 

\usepackage[latin1]{inputenc}  
\usepackage{graphicx}        
\usepackage{bm}              
\usepackage{subfigure}
\usepackage{xcolor}
\usepackage[colorlinks=true,citecolor=blue,urlcolor=blue]{hyperref}

\begin{document}

\title{Field strength scaling in quasi-phase-matching of high-order harmonic generation by low-intensity assisting fields}

\author{Emeric Balogh}
\affiliation{Department of Optics and Quantum Electronics, University of Szeged, 6720 Szeged, Hungary}
\affiliation{ELI-ALPS, ELI-HU Nkft, 6720 Szeged, Hungary}
\affiliation{Center for Relativistic Laser Science, Institute for Basic Science (IBS), Gwangju 500-712, Republic of Korea}
\email{ebalogh@titan.physx.u-szeged.hu}

\author{Katalin Varj\'u}
\affiliation{Department of Optics and Quantum Electronics, University of Szeged, 6720 Szeged, Hungary}
\affiliation{ELI-ALPS, ELI-HU Nkft, 6720 Szeged, Hungary}

\begin{abstract}
High-order harmonic generation in gas targets is a widespread scheme used to produce extreme ultraviolet radiation, however, it has a limited microscopic efficiency.
Macroscopic enhancement of the produced radiation relies on phase-matching, often only achievable in quasi-phase-matching arrangements.
In the present work we numerically study quasi-phase-matching induced by low-intensity assisting fields. 
We investigate the required assisting field strength dependence on the wavelength and intensity of the driving field, harmonic order, trajectory class and period of the assisting field.
We comment on the optimal spatial beam profile of the assisting field.
\end{abstract}

\maketitle

\section{Introduction}

The most promising way of coherent extreme ultraviolet and soft x-ray short pulse generation is high-order harmonic generation (HHG) of near-infrared and mid-infrared laser pulses in gases and solids. 
HHG in gases has the advantage of having less constraints on the laser parameters, however, it also has a moderate conversion efficiency (10$^{-4}$ -- 10$^{-6}$) \cite{2002PRAHergott,2008PQEMidorikawa,2009OLSkantzakis}, which decreases further with the wavelength of the generating laser pulse ($\lambda^{-5.5}$ -- $\lambda^{-6.5}$) \cite{2007PRLTate, 2009PRLShiner}. 
The increase in laser intensity to produce high photon-flux from harmonic radiation is limited by the ionization it produces: through depletion of the medium and distortions of the laser pulse in the plasma. 
The use of long gas cells also raises problems of spatial phase-matching (PM) of the generated harmonic radiation \cite{1995PRLSalieres, 1997PRABalcou, 2008JPBGaarde}. 

In a macroscopic medium phase mismatch arises when the phase velocity of the polarization induced by the propagating laser field is different from that of the propagating harmonic field \cite{1997PRABalcou, 2000RevModPhysBrabec}. 
The exact description of phase-mismatch is a complex task mainly because gas HHG is non-instantaneous: during the process the valence electron leaves the core, gains energy and recombines with the ion releasing its energy in form of a photon. 
Along its trajectory the electron accumulates a phase which is inherited by the harmonic field, making the harmonic's phase dependent on the phase, intensity and shape of the generating laser field as well as the length of the electron trajectory \cite{2008JPBGaarde}.
The complex relation between harmonic phase and generating laser field properties makes PM a complicated process, and there is no known general formula for optimal PM when harmonics are generated in a gas cell or gas jet by a laser pulse producing considerable ionization rate \cite{2008JPBGaarde}. 

The description of phase-matching is greatly simplified when the intensity dependence of the harmonic phase can be ignored, (for ex. in HHG in waveguides, or with loose focusing and low ionization), allowing the construction of a simple analytical model. 
This type of phase-matching has already been discussed extensively \cite{1999PRLDurfee, 2006JQEPaul, 2009PNASPopmintchev, 2010NPhotPopmintchev, 2013rsirudawski}, and now it is known that, depending on the characteristics of the target gas, there is a limit on the achievable photon energy (phase-matching cutoff). 
This limit is imposed by the intensity of the driving field that produces a critical ionization rate above which conventional phase-matching is not possible \cite{2010NPhotPopmintchev}. 

Above this phase-matching limit, quasi-phase-matching (QPM) schemes are often used to increase harmonic yield \cite{2010NPhotPopmintchev}. 
As in gas HHG the traditional QPM schemes based on birefringence are not possible other methods have been proposed. 
These are based on some type of periodic modulation along the propagation axis, which includes 
atomic density (modulated by acoustic waves \cite{2000PRAGeissler}, or using a multijet configuration \cite{2007PRAAuguste,2008NJPTosa,2007NPhysSeres,2011PRLWillner, 2014PRAGaneev}), 
driving field intensity (in modulated waveguides \cite{2000OptExprChristov, 2003NaturePaul, 2003ScienceGibson,2010PRAFaccio}, and by multimode beating in capillaries \cite{2007PRLZepf, 2007OptExprDromey}),  or modulation caused by a secondary periodic field, which is either static \cite{2010PRLBiegert}, or propagating in another direction than the driving field \cite{2007NPhysZhang, 2007JosaBLandreman, 2007PRLCohen, 2007OptLettCohen, 2008OptExprLytle, 2012PRLKatus, 2014OEOKeeffe}.

In this paper we discuss QPM schemes employing secondary assisting fields, and present numerical results that can give an indication to experiments on the intensities of the secondary field required to induce QPM, depending on the parameters of the driving field. 
This paper is organized as follows: In section II we briefly review the known QPM schemes, and summarize their main features, including the optimum phase-shift employed, and the maximum achievable efficiency. 
Starting with section III we describe our new results. 
There the magnitude of the phase-shift produced by the assisting field in terms of the driving field strength, wavelength ratio of the two fields and trajectory class responsible for HHG is presented. 
The presentation of the spatial profile of the optimum assisting field is discussed in section IV, and we conclude in Section V.

\section{Methods of quasi-phase-matching employing low-intensity assisting fields}
\label{sec:II}

QPM is a powerful tool when conventional phase-matching is not possible, i.e. when the phase velocity of the nonlinear polarization created by the driving laser cannot be matched with the phase velocity of the harmonic field, thus phase-mismatch (PMM) arises. 
In the forward propagation direction the magnitude of the wavevector of the propagating harmonic field ($k_h$) and the wavevector of the high-harmonic polarization generated by the laser ($k_H$) are different ($\Delta k = k_H - k_h$).
PMM in gases HHG has four different sources \cite{2013rsirudawski}:
\begin{equation}
\Delta k =  \Delta k_g + \Delta k_n + \Delta k_e + \Delta k_a.
\label{eq:deltak}
\end{equation}
where $\Delta k_g$ arises from the Gouy phase-shift around the focus and $\Delta k_e$ from free electron dispersion. These terms are always negative. 
On the other hand the wavevector mismatch from neutral dispersion ($\Delta k_n$) is always positive. 
The last term ($\Delta k_a$) arises from the intensity dependent atomic phase and it is negative before the focus and positive after the focus. 
In HHG the same harmonic can be generated by electrons performing a short or long trajectory before recombination. 
In case of short trajectories the atomic phase is negligible in many practical cases, and a slowly varying laser intensity (like in loose focusing geometries) can make this contribution negligible for the long trajectories as well.
By neglecting the atomic phase, the above equation shows that phase-matching can only be achieved when the total dispersion contribution $\Delta k_n + \Delta k_e$ is positive, and balances the effect of focusing ($\Delta k_g$).
This relation creates an upper bound to the ionization rate, and limits the maximum laser intensity usable for phase-matched harmonic generation. 
At higher ionization rates PMM is unavoidable \cite{2009PNASPopmintchev}.

\begin{figure*}[htb!]%
	\includegraphics[width=18 cm]{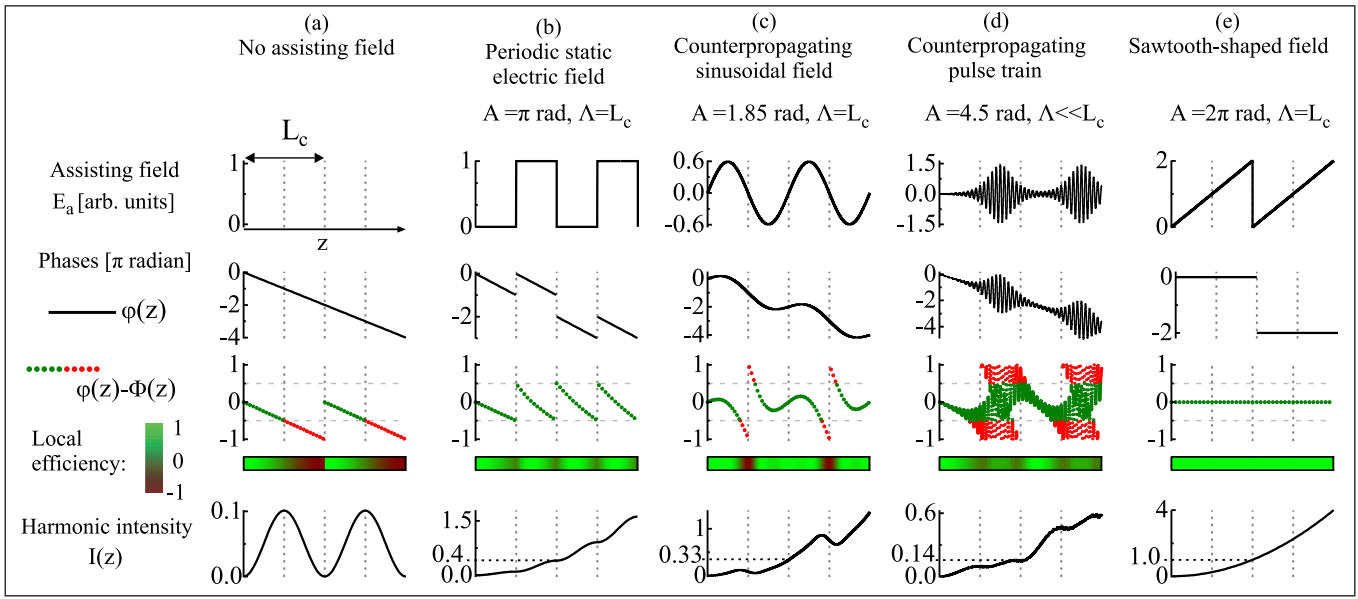}%
\caption{(a) Illustration of phase-mismatch in harmonic generation via phase and intensity variation along the propagation direction $z$. (b)-(e) Schematic presentation of QPM methods employing periodic assisting fields. Top row: illustration of the assisting field distribution in units of $A=\pi E_a$. Second row: Effect of the assisting field on the generated harmonic's phase. Third row: Phase difference (modulo $2\pi$) between the generated and propagated fields and local efficiencies shown in color scale. Bottom row: Harmonic intensities, whose values at $L_{c}$ also show the overall efficiency of the process. $\varphi$ and $\Phi$ denote phases of the generated and the propagated harmonic fields.}%
\label{fig1}%
\end{figure*}

The consequence of PMM is that the intensity of the harmonic field periodically increases and decreases along the propagation axis (\autoref{fig1}.a), which in nonlinear optics is known to be responsible for the appearance of Maker fringes \cite{1962PRLMaker}. 
The harmonic field builds up until the phase difference between the locally generated and the propagated harmonic fields is smaller than $\pi$/2, then, due to destructive interference the harmonic intensity decreases. 
Zones where harmonic intensity increases/decreases are called zones of constructive/destructive interference. 
The basic idea of QPM is to eliminate harmonic emission in destructive zones, or switch these into constructive zones, increasing the harmonic yield over longer propagation distances. 

In multi-jet configuration QPM the elimination of emission in destructive zones is achieved by tuning the gas pressure (and thus the value of $\Delta k_n$ and $\Delta k_e$) so the length of the constructive zone matches the length of a single gas jet and the individual jets are placed at a distance along which vacuum propagation (now only containing $\Delta k_g$ and $\Delta k_a$) continues over the destructive part \cite{2007NPhysSeres}.
By contrast, in the field-assisted configuration the gas cell is continuous and the secondary field is used to shift the phase of the generated harmonics in the destructive zones, turning these into (partially) constructive zones.

In our discussion we assume $\Delta k$ to be constant over the length of the medium, and we neglect the effect of absorption. 
These assumptions are justified in cases when the coherence length is much shorter than the absorption length and the laser intensity is constant along the propagation axis (for example in guided generation, or under loose focusing conditions).
To describe the process it is convenient to use a coordinate frame moving with the phase velocity of the harmonic in question ($z'=z, t'=t-z/v_q$). 
In the following we drop the prime symbol for simplicity.
The phase of the generated harmonic field in the moving frame can be expressed as $\varphi_q (z)=-\Delta k z = -2\pi z/L_{c}$ where L$_{c}$ is the coherence length.

An assisting electric field periodic in space induces a periodic modulation of the polarization phase, so this becomes $\varphi_q (z)=-\Delta k z + A f(z)$, where $A$ is the amplitude of the phase-modulation induced by the assisting field and $f(z)$ is a normalized function with $\Lambda$ spatial periodicity in the moving frame.
QPM methods employing low-intensity assisting fields are based on the fact that the phase-shift induced by the assisting electric field scales linearly with its amplitude ($E_a$) in the limit when that is much weaker than the amplitude of the generating field ($E_a \ll E_0$, see \cite{2007PRLCohen} for details).
As a result, the shape of the phase-modulation resembles that of the assisting field, and its amplitude can be expressed as
\begin{equation}
A=\zeta E_a.
\label{eq:zeta}
\end{equation}
The calculation of $\zeta$ is presented in \autoref{phasemodulation}.

Assuming a normalized emission rate independent of $z$, the near field at the end of cell with length $L$ can be calculated as $H_q(L) \propto \int_0^{L} exp(-i \Delta k z') dz'$, the phase then is given by  $\Phi_q(L)=Arg[H_q(L)]$, while the harmonic intensity will be $I_q(L) \propto  |H_q(L)|^2$.
We define the efficiency of the phase-matching method as
\begin{equation}
\eta = I_q(L)/I^0_q(L),
\label{eq:eta}
\end{equation}
where $I^{0}_q(L)$ is the intensity of the propagated field produced with perfect phase-matching (when $\Delta k=0$).

As seen in \autoref{fig1}.b-e, in QPM the intensity of the generated harmonic increases approximately quadratically with the length of the cell as $\eta \cdot (L/L_c)^2$, with only slight sub-coherence-length oscillations around the parabola. 
The intensity in optimal QPM conditions might increase until it reaches the absorption limit. 
Whereas in the case of PMM \autoref{fig1}.a, the peak intensity is reached at half of the coherence length, severely limiting the achievable photon number in macroscopic media.

Periodic assisting fields that can induce QPM can be of many types: to date periodic static electric fields, perpendicularly propagating THz fields, and counterpropagating (to the IR) quasi-cw laser and sawtooth-shaped fields and pulse trains have been proposed or used. 
Although the experimental realization of the different schemes differ widely, the basic physics behind the phenomena is the same in all cases, and in Section 3 we will show that the optimal amplitude of the assisting field can be calculated with a general formula.

\subsection{Periodic static electric fields}

QPM in HHG by using periodic static electric field has been proposed by Biegert et al. \cite{2010PRLBiegert, 2011NIMASerrat}. 
In this scheme high-order harmonics are periodically generated with no assisting field over half a coherence length, then, just before destructive interference would occur a DC field shifts the phase of the selected harmonic by $A=\pi$, and constructive interference continues over the other half of the coherence length (see \autoref{fig1}.b). 
Alternating zones with and without static electric field create the condition for QPM.
Under the approximations presented earlier, this scheme produces the same efficiency as conventional QPM $(2/(m\pi))^2$ in case of second-harmonic generation, which has been discussed extensively by Fejer et al. \cite{1992JQEFejer}. 
This also means that higher order spatial QPM is possible, where the periodicity is $m\Lambda$, $m$ being a positive integer number. 
For odd $m$ orders the length of 0 and $\pi$ phase-shift zones should be $m L_{c}/2$, however for even orders $(m-1) L_{c}/2$ and $(m+1) L_{c}/2$ long zones should alternate \cite{1992JQEFejer}.
It also follows that higher order QPM in the amplitude of phase-modulation ($n\pi$) is possible, however only for odd $n$ orders, and these produce the same efficiency as first order QPM.
In conclusion, using this scheme 40.5\% efficiency can be obtained by $A=\pi \; rad$  phase-shift with $\Lambda=L_{c}$ periodicity. 

\subsection{Sinusoidal electric fields matching the coherence length}

QPM is also achievable with sinusoidal phase-modulation as illustrated in \autoref{fig1}.c. 
Such schemes were proposed where the phase-modulation is achieved by counterpropagating quasi-cw fields \cite{2007PRLCohen, 2008OptExprRen}, or perpendicularly propagating THz pulses \cite{2012PRLKatus}. 
In both cases the optimal phase-shift induced by the assisting field has to be $A=1.85$ radian (the position of the first extremum of the first order Bessel function of the first kind $J_1(A)$) \cite{2007PRLCohen}. 
Higher order QPMs can be achieved when the spatial period or amplitude of the phase-modulation is higher than required for first order. 
QPM of $m$th order in space and $n$th order in amplitude occurs when the phase-modulation period is $m L_{c}$ and the amplitude is at the position of the $n$th maximum of $(J_m(A))^2$. 
The efficiencies in these cases can be calculated by the values of $(J_m(A))^2$, the highest being 33.7\% for first order QPM \cite{2007PRLCohen}. 

\subsection{Counterpropagating pulse trains}

Another method of QPM is to scramble the phase of the generated harmonics at zones of destructive interference by a counterpropagating pulse or pulse train, that suppresses emission in these regions (see \autoref{fig1}.d).
These schemes have been extensively discussed already \cite{2001PRLVoronov, 2007NPhysZhang, 2007OptLettCohen, 2007JosaBLandreman, 2008OptExprLytle, 2010JosaBRobinson, 2012OptExprKeeffe}, and it has been found that the harmonic emission can be eliminated by counterpropagating light pulse \cite{2004OptExprSutherland, 2007JosaBLandreman} and this can induce QPM \cite{2007NPhysZhang}. 
The intensity of counterpropagating pulse interfering with the forward-propagating driving pulse has to be only a small fraction of the driving intensity to eliminate emission \cite{2007JosaBLandreman}. 
With this method the coherence length should match the width of the counterpropagating pulse, not its wavelength, therefore it is easiest to implement when the coherence length is much larger than the wavelength of the assisting pulse ($\lambda_a \ll L_{c}$) \cite{2007OptLettCohen}. 

For this type of QPM, flat-top laser pulses have been generated and applied experimentally \cite{2010JosaBRobinson,2014OEOKeeffe}, and the effect of $sech^2$-shaped pulses has been analyzed theoretically \cite{2007OptLettCohen}.
Complete elimination of emission in destructive zones can achieve an efficiency of 10.1\% ($1/\pi^2$) with flat-top pulses \cite{2007OptLettCohen}.
However, destructive zones can also be switched into partially constructive zones, increasing efficiency \cite{1997OptExprPeatross, 2007OptLettCohen, 2007NPhysZhang}. 
The phase-shift induced by the counterpropagating light yielding the best efficiency for $sech^2$-shaped pulses is $A=4.5 rad$ (case shown in  \autoref{fig1}.d), increasing the overall efficiency to 14\% \cite{2007OptLettCohen} using the optimal length of the counterpropagating pulse of 0.23 $L_{c}$ (intensity FWHM) \cite{2007OptLettCohen}.
In case of square-shaped pulses the best efficiency of 20\% is produced by a phase-shift of $A=3.83\; rad$, (the global minimum of $J_0(A)$ \cite{2007OptLettCohen}).

The obvious advantage of this scheme is, that any phase-modulation comparable or larger than $\pi$ would produce partial extinction of harmonic yield, therefore this method is not very sensitive to the parameters of the assisting field \cite{2007JosaBLandreman}.

\subsection{Sawtooth-shaped fields}

In theory, perfect elimination of the PMM can be obtained, if a sawtooth-shaped field is applied as proposed in \cite{2010oesidorenko}.
Therefore, this in not a traditional QPM method, we mention it due to the fact that it also uses an assisting field and, in theory, this can achieve 100\% efficiency.

\subsection{Summary}

We conclude that the implementation of all QPM schemes which employ a weak, periodic or quasi-periodic electric field requires a precise determination of the assisting field's parameters to achieve significant enhancement of the macroscopic radiation. 
The assisting field can be described by its period and amplitude.
The periodicity of the field is determined by the coherence length of the high order harmonic generation process which, in some cases, can be calculated \cite{1999PRLDurfee} or even measured \cite{2003PRAKazamias,2007NPhysZhang,2008OptExprLytle}. 
The calculation of the optimal electric field amplitude is presented in the next section.

\section{Magnitude of phase-modulation induced by assisting fields}
\label{phasemodulation}

\subsection{Assisting field wavelength identical with driver wavelength}

\begin{figure}[htb!]%
	\includegraphics[width=8.5 cm]{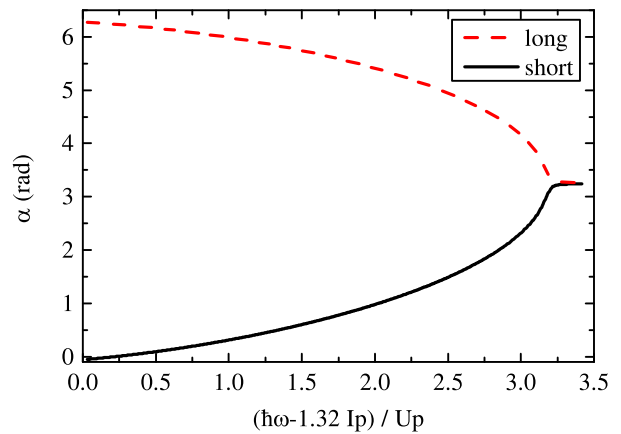}%
\caption{Coefficient of intensity dependent harmonic phase for different trajectories. For more details see \cite{2008JPBGaarde}.}%
\label{fig2}%
\end{figure}

Using assisting fields of the same wavelength as the driver ($\lambda_a = \lambda_0$) the phase-modulation induced by the weak assisting field can be expressed analytically. 
The phase of a harmonic $q$ ($\varphi_q$), generated by a quasi-monochromatic field (apart from a constant) can be expressed as \cite{1997PRABalcou}:
\begin{equation}
  \varphi_{q} = q \varphi_0 - \frac{\alpha U_p}{\hbar \omega_0},
\label{eq:eqphiq}
\end{equation}
where $\varphi_0$ and $U_p$  are the phase  and ponderomotive energy of the generating field, the latter is proportional to the intensity. 
The $\alpha$ coefficient depends on the length of the electron trajectory involved in generating harmonic $q$, and its value can be obtained from classical or quantum mechanical HHG models \cite{2008JPBGaarde}, and it is shown in \autoref{fig2}.

The two terms in the rhs. of \autoref{eq:eqphiq} show that at the interference of two fields the modulation of the harmonic's phase has two sources: the modulation of the driver's phase, and the modulation of the driver's intensity.
Let us call these two the \emph{direct} and \emph{indirect} phase modulations, following the works \cite{2001PRLVoronov} and \cite{2007JosaBLandreman}.
In the limit of $E_a \ll E_0$ both of these contributions will cause the harmonic's phase to change sinusoidally with the phase difference between the driver and assisting field ($\varphi_a - \varphi_0$). 
The amplitude of the direct phase-modulation for harmonic $q$ can be expressed as  (see Appendix for more details):
\begin{equation}
  \Delta\varphi_{p}\approx q\frac{E_a}{E_0}
\label{eq:eqp}
\end{equation}

The indirect phase-modulation (arising from the second term in \autoref{eq:eqphiq}) is linked to the intensity-dependence of the harmonic's phase (i.e. the atomic phase mentioned in Sec.2.).
The amplitude of this modulation can be approximated by
\begin{equation}
  \Delta\varphi_{I}\approx \frac{-\alpha e^2 E_0 E_a}{2 m_e \hbar \omega_0^3}=\frac{-\alpha 2U_p}{\hbar\omega_0}\frac{E_a}{E_0},
\label{eq:eqI}
\end{equation}
where $e$ and $m_e$ denote the electron charge and mass respectively and $\omega_0$ is the angular frequency of the generating laser field (see the Appendix for more details).

\begin{figure}[htb!]%
	\includegraphics[width=8.5 cm]{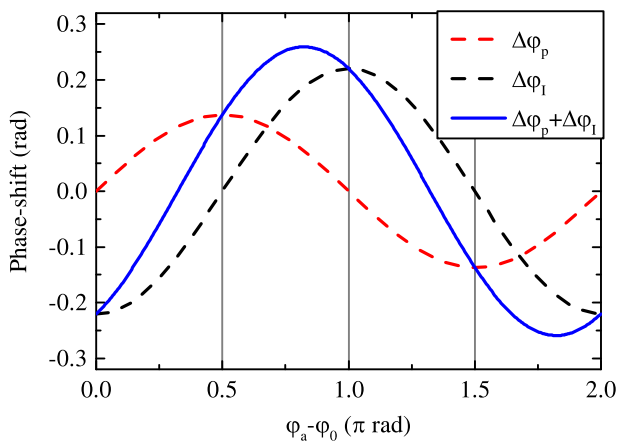}%
\caption{Direct and indirect harmonic phase-modulation caused by interfering driver and assisting waves, shown as a function of the phase difference between the two fields. Phases calculated using \autoref{eq:eqphiq}, with parameters: $\alpha=\pi$, $q=93$, a driver and assisting field with intensities of $6\times10^{14}$ W/cm$^2$ and $1.32\times10^{9}$ W/cm$^2$ respectively, and a wavelength of $\lambda_0=800$ nm. More details in \cite{2001PRLVoronov,2007JosaBLandreman} and Appendix.}%
\label{fig3}%
\end{figure}

The maxima of the two -- direct and indirect -- components of the phase-shift occurs shifted by $\pi / 2$ in phase difference between the two interfering fields, as illustrated in \autoref{fig3}.
Due to this delay, the total phase-modulation can be calculated simply as
\begin{equation}
  A_0 = \sqrt{\Delta\varphi_{I}^2 + \Delta\varphi_{p}^2}.
\label{eq:eqfitot}
\end{equation}
From the above equation, the scaling factor $\zeta$ between the assisting field strength and the phase-shifting effect (of \autoref{eq:zeta}) for cases when the assisting and driver fields have the same wavelength can be expressed as
\begin{equation}
  \zeta_0 = \sqrt{\left( \frac{q}{E_0} \right)^2 + \left( \frac{\alpha 2 U_p}{\hbar \omega_0 E_0} \right)^2}.
\label{eq:eqzetatot}
\end{equation}
For cutoff harmonics it is known that $q = (1.32 I_p + 3.2 U_p)/\hbar\omega_0$ and $\alpha \approx \pi$ (where $I_p>0$ is the ionization energy), therefore in cases when $I_p \ll U_p$, the scaling factor becomes
\begin{equation}
  \zeta_0^{cutoff} \approx \frac{7U_p}{\hbar\omega_0 E_0}\propto E_0 \cdot \lambda_0^{3}.
\label{eq:eqfitot3}
\end{equation}
This shows that for cutoff harmonics, where QPM methods are found to be most effective, the required field strength scales inversely with the driving field strength and the third power of its wavelength:
\begin{equation}
  E_a^{cutoff} = \frac{A}{\zeta^{cutoff}}\propto E_0^{-1}\lambda_0^{-3}.
\label{eq:eqfitot4}
\end{equation}
As the cutoff energy in HHG scales as $E_0^2 \cdot \lambda_0^2$, the same energy photons still require weaker assisting fields when generated by weaker, but longer wavelength driver fields.

\subsection{Assisting IR field with different wavelength}

With an assisting field of arbitrary wavelength we rely on numerical calculations to obtain the same information. 
We use the nonadiabatic saddle-point approximation to calculate the harmonic phases \cite{2004PRASansone,1994PRALew}. 

Saddle-points of the Lewenstein integral are known to reproduce well the phase-derivatives of the generated harmonics.
Using this method the phase of a selected harmonic can be expressed as $\varphi_q = q\omega_0 t_r - S(t_i,t_r)/\hbar$, where $t_i$ and $t_r$ are the solutions of the saddle-point equations representing the ionization and return times of the most relevant electron trajectories, and $S(t_i,t_r)$ is the quasiclassical action.
The solved equations read as:
\begin{align}
\textbf{p}_s=\frac{1}{t_r-t_i}\int_{t_i}^{t_r} \textbf{A}(t) dt \\
\frac{[\textbf{p}_s+\textbf{A}(t_i)]^2}{2}-I_p=0 \\
\frac{[\textbf{p}_s+\textbf{A}(t_r)]^2}{2}+I_p=q \hbar \omega_0 
\label{eq:saddlep}
\end{align}
where $\textbf{p}_s$ is the stationary value of the canonical momentum and $\textbf{A}(t)$ is the vector potential which has no direct relation to the scalar $A$ used in the other equations.

To calculate the effect of the assisting field, we calculate electron trajectories in the two-color field while changing the phase-difference between the two fields.
From this the oscillating harmonic phase $\varphi_q$ like the blue solid line in Fig3 can be obtained, revealing the amplitude of the total phase-modulation.

\begin{figure}[htb!]%
	\includegraphics[width=8.5 cm]{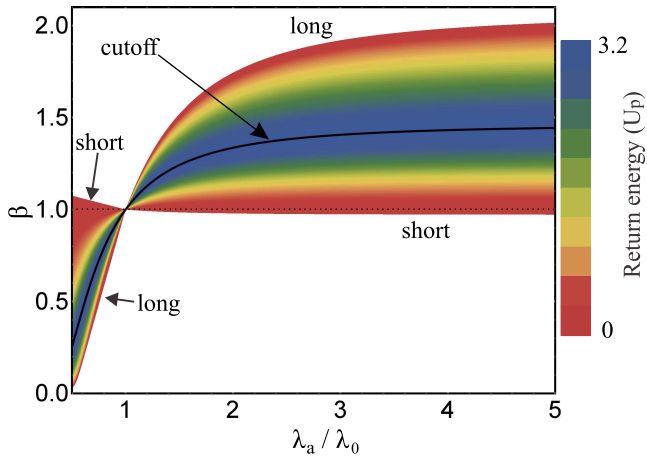}%
\caption{Phase-modulation coefficients for arbitrary wavelength assisting fields. Trajectory lengths ($\tau$) are shown in color scale represented by their final kinetic energy in units of $U_p$.}%
\label{fig4}%
\end{figure}

We observe that for the obtained phase-modulation amplitude the same scaling rules apply than in the previous case.
In fact  the obtained phase-modulation effect ($\zeta$) can be related to the previous case where the two fields had the same wavelength (characterized by $\zeta_0$) and a simple correction factor can be introduced:
\begin{equation}
\zeta = \zeta_0 \beta(\lambda_a / \lambda_0,\tau)
\label{eq:beta}
\end{equation}
where $\tau$ distinguishes trajectories with different travel times, and $\lambda_a / \lambda_0$ is the ratio of the two wavelengths. 
The value of $\beta(\lambda_a / \lambda_0,tr)$ obtained from the saddle-point solutions is shown in \autoref{fig4}. 

\begin{figure}[htb!]%
	\includegraphics[width=8.5 cm]{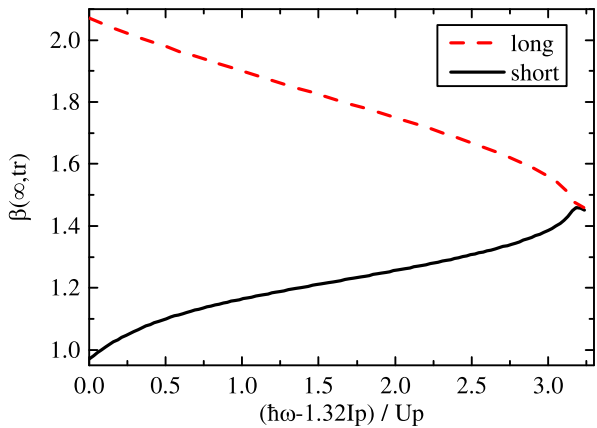}%
\caption{Coefficient for calculating phase-modulation caused by static electric field. Trajectories represented by their final kinetic energy in units of $U_p$.}%
\label{fig5}%
\end{figure}

By definition of the parameters, at $\lambda_{a} = \lambda_{0}$ all the trajectory dependence of the phase shift is included in $\zeta_0$. 
It is interesting to see how the correction factor for short and long trajectory components cross at this point.
For all values of $\lambda_{a} > \lambda_{0}$ we observe that the value of the correction factor is almost constantly 1 for the shortest trajectories, and for the longest trajectories the deviation from 1 has the largest magnitude. 
This finding is consistent with the simple view, that the longer the electron stays in the continuum, the more sensitive it becomes to the effect of the assisting field \cite{2007PRLCohen}.
In the limit when $\lambda_{a} \gg \lambda_{0}$ (i.e. when the assisting field can be considered static during the electron's travel in the continuum), the value of the $\beta$ correction factor goes to 1.45 for cutoff harmonics.
For other trajectories this factor varies as shown in \autoref{fig5}, being very close to one in case of the shortest trajectories and going slightly higher than two for trajectories with a return time of one optical cycle. 
For assisting fields with $\lambda_a > 5\lambda_0$ the values of $\beta$ calculated for static fields (shown in \autoref{fig5}) are already reasonably accurate.

The figure also indicates, that for $\lambda_{a} < \lambda_{0}$ the dependence of the correction factor on trajectory length is reversed; the longest the trajectory, the less the effect -- which can be understood as the perturbation caused by the assisting field can average out through the longer traveling time of the electron. 
This means that in this regime the relative effect of the assisting field on shorter trajectories becomes more and more pronounced. 
We would like to point out the practicality of this limit: since the assisting field's wavelength is determined by the coherence length, and $L_c$ scales inversely with harmonic order \cite{2000PRAGeissler}, it might reach very small values when increasing driver wavelengths are applied to generate very high harmonics in the x-ray region \cite{2008PRLTakahashi, 2012spopmintchev}. 
In this scenario under very unfavorable PM conditions short wavelength assisting fields might be useful in achieving QPM.

We performed calculations with different laser field and ionization potential parameters, all yielding very similar results to what is shown in \autoref{fig4}, only finding small deviations from it. 
The results are found to be more accurate in the high-intensity regime, where $U_p > I_p$.

Finally, combining equations \ref{eq:zeta}, \ref{eq:eqzetatot} and \ref{eq:beta}, the formula for the strength of the assisting field causing the required $A$ phase-modulation for harmonic order $q$ can be expressed as
\begin{equation}
E_a = \frac{A E_0}{\beta \sqrt{q^2+\left(  \frac{2 \alpha U_p}{\hbar \omega_0}  \right)^2 }}
\label{eq:ea}
\end{equation}
the value of $\beta$ depending on the ratio of the driver and assisting fields wavelength, and shown in \autoref{fig4}.

We note here, that in some QPM schemes the wavelength of the assisting field, $\lambda_a$ is not a free choice, it is determined by the coherence length. 
This implies, that the correction factor $\beta$ has only an indirect dependence on the generating laser pulse parameters through the coherence length, and depends directly only on the chosen trajectory, thus the scaling law expressed in \autoref{eq:eqfitot4} for cutoff harmonics holds generally.

In many practical cases, especially in free focusing geometries, the intensity of the beam can change along the propagation axis, moreover the pulse shape and structure is also affected by dispersion, self-phase modulation and diffraction caused by non-linear effects. 
These effects also change the coherence length along the propagation axis. 
To compensate this effect the assisting field's periodicity has to match the changing $L_c$ along the whole gas medium for efficient QPM. 
To this end, the use of chirped assisting fields \cite{2007PRLCohen,2012PRLKatus}, or counter-propagating pulse-trains with variable separation has been proposed \cite{2010JosaBRobinson}. 
When using chirped assisting pulses, however, the wavelength ratios become ill-defined. In these cases our results remain indicative. 

Another important aspect to mention is that the coherence length is also time-dependent: the ionization rate is increasing within the duration of the driver pulse. For the best overall efficiency it is desired that phase-matching or QPM is achieved around the peak of the pulse, where the microscopic generation efficiency is usually highest, and where the highest harmonics are generated.

Finally, we note that our model is obviously not applicable to the cases when the assisting field alone causes photo-ionization, replacing the role of tunneling ionization in the three-step model of HHG.

\section{Assisting beam profile}

In order to induce QPM in a macroscopic media the phase-shifting effect of the assisting field for a given harmonic has to be the same at different spatial coordinates across the beam.
However, due to the intensity profile across the beam the same harmonic order falls at different parts of the plateau and thus has an $\alpha$ and $\beta$ value varying with the radial coordinate.
Both of these affect the phase-shifting effect of the assisting field. 
To compensate this, the assisting field must have an appropriate spatial profile.

\begin{figure}[htb!]%
	\includegraphics{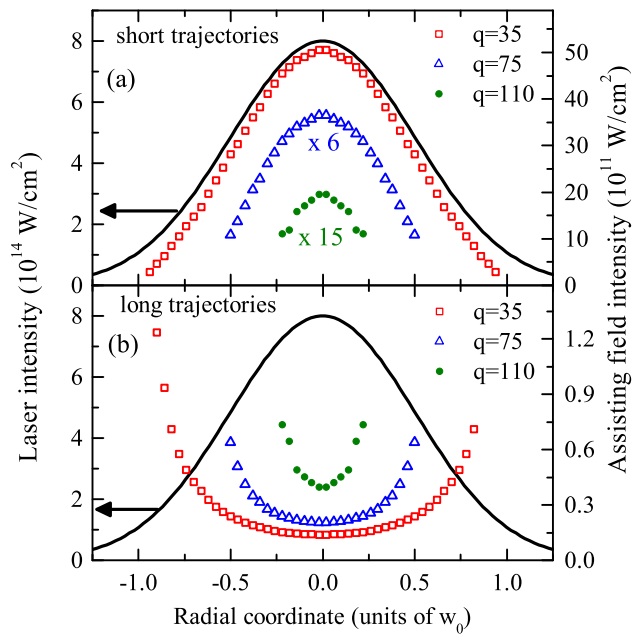}%
	\caption{Numerically calculated radial intensity profiles needed to induce $\pi$ phase-shift for different plateau harmonics, when generated by a Gaussian beam having a beam radius of w$_0$, and peak intensity $8 \times 10^{14}$ W/cm$^2$. Calculations done for $\lambda_0 = 800$ nm driving field, and long wavelength assisting fields $\lambda_a \gg \lambda_0$. At the peak intensity the values of $(\hbar \omega - 1.32 I_p)/U_p$ for harmonics 110, 75 and 35 are 2.96, 1.8 and 0.54 respectively.}%
\label{fig6}%
\end{figure}

Assuming that the generating laser beam has a Gaussian spatial profile, the intensity profile of the assisting field can be determined using numerical calculations presented in the previous section. 
In \autoref{fig6} the calculated intensity profile is shown for different harmonics generated by 800 nm driving field for the case when A=$\pi$ radian, and $\lambda_a \gg \lambda_0$.

In case of short trajectories a lower IR intensity (off-axis) means that the same harmonic is closer to the cutoff, has both higher $\alpha$ and higher $\beta$ values, therefore the required assisting field strength is lower. 
The opposite stands for long trajectories, where $\alpha$ and $\beta$ decreases with intensity (see \autoref{fig2}). 
This issue is not risen for cutoff harmonics which are only generated close to the axis. 

The intensity profile required by QPM (\autoref{fig6}) for short trajectories closely resembles a Gaussian suggesting that counter-propagating fields may be used to induce QPM in the whole cross section of the gas cell. 
As for long trajectories the required field intensity is higher off-axis, which could be an explanation why the most efficient QPM was found for harmonics close to the cutoff \cite{2010PRLBiegert, 2012PRLKatus}. 
Another important aspect of \autoref{fig6} is that for long trajectories the required field strength is two orders of magnitude lower and  almost constant for different harmonics (close to the axis), while for short trajectories it shows high variation with harmonic order (a result consistent with the findings of Zhang et al. \cite{2007NPhysZhang}). 
Thus spectral selection might be easier to achieve for short trajectories by varying the strength of the assisting field.

However, in case of counter-propagating pulse trains, the only constraint for (partial) elimination of harmonic emission from destructive zones is that the phase-shift should be larger than $\pi$.
In this respect short trajectories dominate the selection of the field strength, since those always require higher intensity assisting field for the same phase-shift. 
This is also consistent with the findings of Landreman et al. \cite{2007JosaBLandreman}.

It should be noted that across the generating beam, not only the driver intensity, but also the coherence length can vary, which can limit the efficiency of QPM, even when using an assisting field with optimal beam profile.

\section{Conclusions}
\label{sec:conclusions}

In this paper we first reviewed proposed QPM techniques employing periodic assisting fields. In ideal situations these methods can yield an enhancement of the harmonic signal along the cell in contrast to the oscillating output from the phase-mismatched situation. The enhancement achievable is of the order of 
$(0.14-0.4)\cdot (\frac{L}{L_c})^2$ in contrast to the oscillation between 0 to 0.1 for the PMM case. We
discussed the required field strength of the assisting field for the implementation of efficient QPM. The presented formula -- Eq. \ref{eq:ea} together with \autoref{fig4} -- can be used to determine its value for experimental realization. 

In conclusion we have analyzed how the phase of high-order harmonic radiation generated by an infrared laser field can be manipulated by low-intensity assisting fields in order to achieve quasi-phase-matching.
A general formula was presented that allows the calculation of the optimal assisting field strength in terms of the generating laser pulse intensity, on the two fields' relative wavelength and the length of the trajectory in question.
We discussed the relationship between the simplest case of counter-propagating assisting fields with the same wavelength (that is analytically treatable), to the case when a different wavelength assisting field is used, and showed that the two can be related through a wavelength-dependent correction factor.
The optimal field profile of assisting fields for short and long trajectory components required for efficient QPM was also discussed, and found that short trajectories have the advantage of requiring the same profile for the driver and the assisting beam.

\textbf{Funding.} 

Hungarian Scientific Research Fund (OTKA NN107235).
K.V. acknowledges the support of the Bolyai Foundation.

\textbf{Acknowledgments.} 
We thank the NIIF Institute for computation time.

\appendix
\section{Harmonic phase modulation in interfering laser fields}
\label{appendix}

Here we describe the harmonic phase modulation induced by the interference of the driver with a weak assisting field, resulting \autoref{eq:eqp} and \autoref{eq:eqI}.
Let us take a driver laser field in the HHG medium along the $z$ axis, described as $E_0 sin (\varphi_0)$, where $\varphi_0 = \omega_0 t + |\vec{k}_0 \cdot \vec{z} |$ is the phase of the laser field, which is a function of space and time.
The assisting field can be described as $E_a sin (\varphi_0 + \delta)$, where $\delta = \varphi_a - \varphi_0$ denotes the phase difference between the two fields, and $\varphi_a = \omega_0 t + |\vec{k}_a \cdot \vec{z}|$ is the phase of the low-intensity assisting field.
Thus $\delta$ becomes dependent on $z$, when the wavevectors $\vec{k}_0$ and $\vec{k}_a$ enclose a nonzero angle.
The resulting wave is
\begin{equation}
E_{tot} sin(\varphi_0+\theta_{tot}) = E_0 sin(\varphi_0) + E_a sin(\varphi_0+\delta),
\label{eq:interference}
\end{equation}
with $E_{tot}$ denoting its amplitude and $\theta_{tot}$ its phase.

In the following calculations we use that $E_a \ll E_0$.
The phase of the resulting wave then can be expressed as
\begin{align}
\theta_{tot} &= arctan\left[\frac{E_a sin(\delta)}{E_0+E_a cos(\delta)}\right] = arctan\left[\frac{\frac{E_a}{E_0} sin(\delta)}{1+\frac{E_a}{E_0} cos(\delta)}\right] \nonumber \\
			       &\approx arctan\left[\frac{E_a}{E_0} sin(\delta)\right] \approx \frac{E_a}{E_0} sin(\delta).
\label{eq:direct}
\end{align}
This term has its first maximum at $\delta = \pi/2$, and the magnitude of the phase-modulation is given by $E_a/E_0$.

In HHG the phase of the generated harmonic depends on the phase of the generating wave multiplied by the harmonic order $q$ \cite{1997PRABalcou}.
As a result the phase-modulation of the generating field described by \autoref{eq:direct} will translate to a \emph{direct} modulation of the harmonic phase with amplitude
\begin{equation}
\Delta \varphi_p \approx q \frac{E_a}{E_0}.
\end{equation}

The harmonic's phase also depends on the intensity of the generating wave, this contribution is usually referred as the atomic (or dipole) phase, because it is inherited from the electron which accumulates it during its travel from ionization to recombination.
This is well approximated as $\varphi_I = -\alpha U_p / (\hbar \omega_0)$, where $U_p = e^2 E^2 / (4 m_e \omega_0^2)$ is the ponderomotive energy in the driver field \cite{2008JPBGaarde}, and the value of $\alpha$ is shown \autoref{fig2}. 
The amplitude of the resulting wave in \autoref{eq:interference} is given by
\begin{equation}
E_{tot} = \sqrt{E_0^2+E_a^2+2 E_0 E_a cos(\delta)},
\label{eq:intens}
\end{equation}
therefore the generating field's amplitude is also modulated with changing $\delta$ and it has its first maximum at $\delta=0$.
This amplitude modulation causes an \emph{indirect} modulation of the harmonic phase. 
In this case the atomic phase is expressed as
\begin{equation}
\varphi_I = \frac{-\alpha e^2}{4 m_e \hbar \omega_0^3}\left[ E_0^2 + E_a^2 + 2 E_0 E_a cos(\delta) \right],
\label{eq:indirect}
\end{equation}
which is modulated with an amplitude given by
\begin{equation}
\Delta \varphi_I = \frac{-\alpha e^2}{2 m_e \hbar \omega_0^3} E_0 E_a.
\end{equation}

Both of these factors have been described in \cite{2001PRLVoronov,2007JosaBLandreman}, for the most relevant case, when the phase difference is calculated for two counterpropagating waves, where $\varphi_0 = \omega t - k_0 z$ and $\varphi_a = \omega t + k_0 z$, giving $\delta = 2 k_0 z$. 

These two different sources of phase-modulation are comparable in magnitude, but the description above shows that they always occur with a shift of $\pi/2$ in the value of $\delta$.
This relation forms the basis of our discussion in \autoref{phasemodulation}. 

\bibliography{qpm}


\end{document}